\documentclass{article}
\usepackage{amsmath}
\usepackage{amsfonts}
\usepackage{makeidx}
\usepackage{cite}
\usepackage{hyperref}
\usepackage{graphicx}
\usepackage{caption}

\newtheorem{theorem}{Theorem}

\newtheorem{example}[theorem]{Example}

\input{tcilatex}
\begin{document}

\title{On Entanglement Measures: Discrete Phase Space and Inverter-Chain
Link Viewpoint}
\author{Felix A. Buot \\
%EndAName
C\&LB Reseaarch Institute, Carmen, Cebu 6005, Philippines;\\
LCFMNN, TCSE Group, Department of Physics, \\
University of San Carlos, Talamban, Cebu City 6000, Philippines}
\maketitle

\begin{abstract}
In contrast to abstract statistical analyses in the literature, we present a
concrete physical diagrammatic model of entanglement characterization and
measure with its underlying discrete phase-space physics. This paper serves
as a pedagogical treatment of this complex subject of entanglement measures.
We review the important inherent concurrence property of entangled qubits,
as well as underscore its emergent qubit behavior. From the discrete phase
space point of view, concurrence translates to translation symmetry of
entangled binary systems in some quantitative measure of entanglement.
Although the focus is on bipartite system, the notion is readily extendable
to multi-partite system of qubits, as can easily be deduced from the
physical inverter-chain link model. A diagrammatic analysis of the
entanglement of formation for any multi-partite qubit system is given.
\end{abstract}

\section{Introduction}

Quantum entanglement has developed from a mere intellectual curiosity \cite%
{EPR} of the fundamental structure of quantum mechanics\footnote{%
Note that although Bell's theorem asserts the nonlocality of quantum
mechanics, the EPR inquiry is still not resolved, i.e., what is is still
left unanswered is the mysterious 'link' between qubits corresponding to our
"see-saw" or mechanical inverter-chain link.
\par
(Note: Bell's inequality theorem is widely discussed in the literature and
websites, we prefer not to cite specific reference).} to become an important
and practical resource for quantum information processing in the evolving
theory of quantum information and ultra-fast computing. Thus, the
quantitative measure of entanglement has developed into one of the most
active fields of theoretical and experimental research. Here we will try to
shed more light on some of the important concepts in the quantification of
quantum entanglement by using a concrete simple mechanical model of a
bipartite system of qubits or chain of qubits. This treatment will be in
contrast with mostly abstract and statistical treatment of entanglement
measure in the literature.

For the two-qubit entanglement, we will focus on the so-called entanglement
of formation and concurrence, two of the most important concepts to
characterize entanglement resource. Here we consider a qubit as a two-state
system. Moreover, we also consider an entangled qubit as effectively a
two-state system, an \textit{emergent qubit} as depicted in our physical
inverter-chain link model. A two-state system has a unity entropy. Thus, a
maximally entangled state has entropy equal to $1$. From the discrete phase
space point of view, any two-state system can be considered to possess two
lattice-site states. Discrete Fourier transform or Hadamard transform
implements unitary superposition of the two \textit{lattice-site states}
(also referred to as '\textit{Wannier functions'}) to yield a sort of 
\textit{crystal-momentum states} (also referred to as '\textit{Bloch
functions'}). For example, take the $\left\vert 00\right\rangle $ and $%
\left\vert 11\right\rangle $ '\textit{lattice-site states}', then the
Hadamard bijective discrete transformation gives the '\textit{%
crystal-momentum states}', $\Phi ^{+}$ and $\Phi ^{-}$, which are two of the
Bell basis states (or '\textit{Bloch function'} states).

\section{Bell Basis Deduced from Inverter-Chain Link Model}

\begin{figure}[h]
\centering
\includegraphics[width=5.0246in]{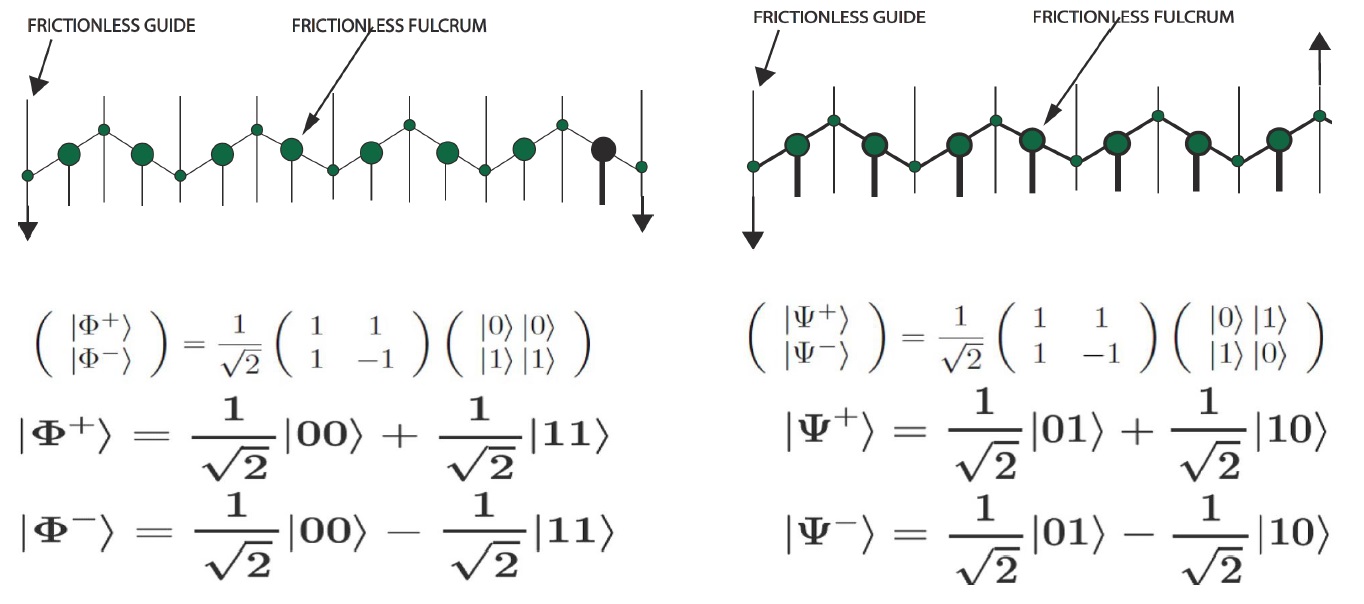}
\caption{Physical diagrammatic model of "triplet" (left) and singlet (right)
entanglement. By construction, each diagram is viewed as a two-state system,
respectively. The Bell basis are readily derived below each diagram, using
the Hadamard transformation. The actual physical implementation of the chain
of inverters may need frictionless male/female sliding tube coupling for
large-angle swing, but this is beside the point.}
\label{fig1}
\end{figure}

We sketch here the derivation of the Bell basis states from our physical
model, as depicted in Fig. \ref{fig1}. The derivation is based in
considering each entangled diagram as a two-state (binary system) of \textit{%
emergent} qubit. Clearly, the entanglement of two \textit{bare} qubits is
divided into two orthogonal spaces of triplet \footnote{%
The use of the term "triplet" is actually a misnomer here since the
entangled system is not free to assume a singlet or zero spin state. Thus,
this term is used here only as a label.} and singlet entanglement states.

We also have the inverse relationship%
\begin{equation}
\left( 
\begin{array}{c}
\left\vert 00\right\rangle \\ 
\left\vert 11\right\rangle%
\end{array}%
\right) =\frac{1}{\sqrt{2}}\left( 
\begin{array}{cc}
1 & 1 \\ 
1 & -1%
\end{array}%
\right) \left( 
\begin{array}{c}
\left\vert \Phi ^{+}\right\rangle \\ 
\left\vert \Phi ^{-}\right\rangle%
\end{array}%
\right)  \label{eqa}
\end{equation}%
\begin{equation}
\left( 
\begin{array}{c}
\left\vert 01\right\rangle \\ 
\left\vert 10\right\rangle%
\end{array}%
\right) =\frac{1}{\sqrt{2}}\left( 
\begin{array}{cc}
1 & 1 \\ 
1 & -1%
\end{array}%
\right) \left( 
\begin{array}{c}
\left\vert \Psi ^{+}\right\rangle \\ 
\left\vert \Psi ^{-}\right\rangle%
\end{array}%
\right)  \label{eqb}
\end{equation}%
We refer to $\left\{ \left\vert 00\right\rangle ,\left\vert 11\right\rangle
\right\} $ as the "\textit{Wannier functions}" space and the $\left\{
\left\vert \Phi ^{+}\right\rangle ,\left\vert \Phi ^{-}\right\rangle
\right\} $ as the corresponding "\textit{Bloch functions}" space of a
two-state \textit{triplet} entanglement system. Likewise, $\left\{
\left\vert 01\right\rangle ,\left\vert 10\right\rangle \right\} $ as the "%
\textit{Wannier functions}" space and the $\left\{ \left\vert \Psi
^{+}\right\rangle ,\left\vert \Psi ^{-}\right\rangle \right\} $ as the
corresponding "\textit{Bloch functions}" space of a two-state singlet
entanglement system.

By virtue of this bijective relationship, any function of Wannier functions
will have a corresponding function of Bloch functions. For example, a
maximally \textit{mixed} Wannier function states will generate a maximally
mixed Bloch function states or the so-called maximally \textit{mixed}
entanglement states. Mixed and pure states will be further discussed below.

\section{ Entangled Qubits as an Emergent Qubit}

The virtue of our inverter-chain link model is that the emergent two-state
property of entangled qubits is very transparent, since changing the state
of one of the qubits also \textit{immediately} change the states of the rest
of the entangled partner(s) \footnote{%
The idea that entanglement is due to conservation of momentum does not hold
for triplet entanglement since its two states give opposing spin-angular
momentum. However, one may interpret that quantum superposition of the
two-opposing angular momentum states conserves the overall zero net
spin-angular momentum, as supported by their eigenvalues similar to a single
qubit. On the other hand, for singlet entanglement, the zero angular
momentum is conserve in both two states. This seemingly apparent physical
difference of the triplet and singlet entanglements underscores the
importance of resolving the "mysterious link" in the EPR inquiry, \cite{EPR}
in order to further advance theoretical physics.}. Here, we will rigorously
justify our claim that entangled qubits behave, as a whole, as an \textit{%
emergent} qubit.

Here, we employ the matrix representation of states and operators. We now
represent $\left\vert \Phi ^{+}\right\rangle $ and as%
\begin{eqnarray*}
\left\vert \Phi ^{+}\right\rangle &=&\frac{1}{\sqrt{2}}\left( 
\begin{array}{c}
1 \\ 
1%
\end{array}%
\right) \\
\left\vert \Phi ^{-}\right\rangle &=&\frac{1}{\sqrt{2}}\left( 
\begin{array}{c}
1 \\ 
-1%
\end{array}%
\right)
\end{eqnarray*}%
Then we have%
\begin{eqnarray*}
\left\langle \Phi ^{+}\right\vert \sigma _{x}\left\vert \Phi
^{+}\right\rangle &=&\frac{1}{2}\left( 
\begin{array}{cc}
1 & 1%
\end{array}%
\right) \left( 
\begin{array}{cc}
0 & 1 \\ 
1 & 0%
\end{array}%
\right) \left( 
\begin{array}{c}
1 \\ 
1%
\end{array}%
\right) =1 \\
\left\langle \Phi ^{-}\right\vert \sigma _{x}\left\vert \Phi
^{-}\right\rangle &=&\frac{1}{2}\left( 
\begin{array}{cc}
1 & -1%
\end{array}%
\right) \left( 
\begin{array}{cc}
0 & 1 \\ 
1 & 0%
\end{array}%
\right) \left( 
\begin{array}{c}
1 \\ 
-1%
\end{array}%
\right) =-1
\end{eqnarray*}%
proving the two eigenvalues, like a qubit property of the triplet system.

For the singlet, it is more convenient to change the phase of the second
term by $i$ so that we have%
\begin{eqnarray*}
\left\vert \Psi ^{+}\right\rangle &=&\frac{1}{\sqrt{2}}\left( 
\begin{array}{c}
1 \\ 
i%
\end{array}%
\right) \\
\left\vert \Psi ^{-}\right\rangle &=&\frac{1}{\sqrt{2}}\left( 
\begin{array}{c}
1 \\ 
-i%
\end{array}%
\right)
\end{eqnarray*}%
Then, we have%
\begin{eqnarray*}
\left\langle \Psi ^{+}\right\vert \sigma _{y}\left\vert \Psi
^{+}\right\rangle &=&\frac{1}{2}\left( 
\begin{array}{cc}
1 & -i%
\end{array}%
\right) \left( 
\begin{array}{cc}
0 & -i \\ 
i & 0%
\end{array}%
\right) \left( 
\begin{array}{c}
1 \\ 
i%
\end{array}%
\right) =1 \\
\left\langle \Psi ^{-}\right\vert \sigma _{y}\left\vert \Psi
^{-}\right\rangle &=&\frac{1}{2}\left( 
\begin{array}{cc}
1 & i%
\end{array}%
\right) \left( 
\begin{array}{cc}
0 & -i \\ 
i & 0%
\end{array}%
\right) \left( 
\begin{array}{c}
1 \\ 
-i%
\end{array}%
\right) =-1
\end{eqnarray*}%
again proving the qubit property of the singlet system, where the two-state
aspect which is quite apparent in Fig. (\ref{fig1}).

The above manipulation for the singlet is a bit round-about trick for the
singlet entanglement. The most straightforward manipulation is to recognize
that the singlet system as an independent system with its own two states
which is orthogonal to the triplet states. In this sense we can also
represent the singlet states just like that of the triplet states, namely,%
\begin{eqnarray*}
\left\vert \Psi ^{+}\right\rangle &=&\frac{1}{\sqrt{2}}\left( 
\begin{array}{c}
1 \\ 
1%
\end{array}%
\right) \\
\left\vert \Psi ^{-}\right\rangle &=&\frac{1}{\sqrt{2}}\left( 
\begin{array}{c}
1 \\ 
-1%
\end{array}%
\right)
\end{eqnarray*}%
Note that in the 'Bloch-state" space, $\sigma _{z}$ in the "Wannier-state"
space is transfrom to $\sigma _{x}$ in entangled states,%
\begin{eqnarray*}
\sigma _{x} &=&H\sigma _{z}H \\
&=&\frac{1}{\sqrt{2}}\left( 
\begin{array}{cc}
1 & 1 \\ 
1 & -1%
\end{array}%
\right) \left( 
\begin{array}{cc}
1 & 0 \\ 
0 & -1%
\end{array}%
\right) \frac{1}{\sqrt{2}}\left( 
\begin{array}{cc}
1 & 1 \\ 
1 & -1%
\end{array}%
\right) \\
&=&\left( 
\begin{array}{cc}
0 & 1 \\ 
1 & 0%
\end{array}%
\right)
\end{eqnarray*}%
Thus, we also have for the singlet states%
\begin{eqnarray}
\left\langle \Psi ^{+}\right\vert \sigma _{x}\left\vert \Psi
^{+}\right\rangle &=&\frac{1}{2}\left( 
\begin{array}{cc}
1 & 1%
\end{array}%
\right) \left( 
\begin{array}{cc}
0 & 1 \\ 
1 & 0%
\end{array}%
\right) \left( 
\begin{array}{c}
1 \\ 
1%
\end{array}%
\right) =1  \label{singlet1} \\
\left\langle \Psi ^{-}\right\vert \sigma _{x}\left\vert \Psi
^{-}\right\rangle &=&\frac{1}{2}\left( 
\begin{array}{cc}
1 & -1%
\end{array}%
\right) \left( 
\begin{array}{cc}
0 & 1 \\ 
1 & 0%
\end{array}%
\right) \left( 
\begin{array}{c}
1 \\ 
-1%
\end{array}%
\right) =-1  \label{singlet2}
\end{eqnarray}%
Following this notion, Eqs. (\ref{singlet1}) and (\ref{singlet2}) can easily
be extended to all entangled qubits, bipartite or multi-partite qubit
systems to yield an emergent qubit. It is far more simpler to analyze the
inverter-chain link diagrams or employ diagrammatic analyses.

\section{Translational/Shift Invariance of Bell Basis}

The translational invariance of maximally entangled Bell basis states
demonstrates a unique characteristic of entangled qubits. This unique
property has been used to detect or measure how much entanglement is present
in arbitrary pure and mixed states. This is implemented in terms of the
notion of concurrence, to be elaborated below.

The translational or shift operation (also rreferred to as the \textit{flip}
operation in the literature) is demonstrated here for the Bell Basis states
and their superpositions. First, let us consider all the Bell basis states.
We have%
\begin{equation*}
\Phi ^{+}=\frac{1}{\sqrt{2}}\left( \left\vert 0\right\rangle \left\vert
0\right\rangle +\left\vert 1\right\rangle \left\vert 1\right\rangle \right)
\end{equation*}%
Upon applying the translation (shift by $+1\left( \func{mod}2\right) $)
operator, $T\left( +1\right) $, we obtain%
\begin{eqnarray*}
T\left( +1\right) \Phi ^{+} &=&\frac{1}{\sqrt{2}}\left( \left\vert
0+1\right\rangle \left\vert 0+1\right\rangle +\left\vert 1+1\right\rangle
\left\vert 1+1\right\rangle \right) \\
&=&\frac{1}{\sqrt{2}}\left( \left\vert 1\right\rangle \left\vert
1\right\rangle +\left\vert 0\right\rangle \left\vert 0\right\rangle \right)
\\
&=&\Phi ^{+}
\end{eqnarray*}%
where addition obeys modular arithmetic ($mod$ $2$). Similarly, we have%
\begin{eqnarray*}
T\left( +1\right) \Phi ^{-} &=&-\Phi ^{-} \\
T\left( +1\right) \Psi ^{-} &=&-\Psi ^{-} \\
T\left( +1\right) \Psi ^{+} &=&\Psi ^{+}
\end{eqnarray*}%
We still consider the $-\Phi ^{-}$ and $-\Psi ^{-}$as invariant since it
differs from the unshifted Bell states by a global phase factor. In the
literature this translation operation is the so-called flipping operation
first used by Wooters, et al, \cite{wooters1, wooters2}

\subsection{Translational property of maximal superposition of Bell basis}

Now let us consider the superposition of maximally entangled Bell basis
states. We can readily see that applying the Hadamard transformation to two 
\textit{Bloch} states will yield the \textit{Wannier} states. We have%
\begin{equation}
\frac{1}{\sqrt{2}}\left( 
\begin{array}{cc}
1 & 1 \\ 
1 & -1%
\end{array}%
\right) \left( 
\begin{array}{c}
\Phi ^{+} \\ 
\Phi ^{-}%
\end{array}%
\right) =\left( 
\begin{array}{c}
\left\vert 0\right\rangle \left\vert 0\right\rangle \\ 
\left\vert 1\right\rangle \left\vert 1\right\rangle%
\end{array}%
\right)  \label{eq1}
\end{equation}%
Thus, 
\begin{eqnarray}
\frac{1}{\sqrt{2}}\left( \Phi ^{+}+\Phi ^{-}\right) &=&\left\vert
0\right\rangle \left\vert 0\right\rangle  \notag \\
\frac{1}{\sqrt{2}}\left( \Phi ^{+}-\Phi ^{-}\right) &=&\left\vert
1\right\rangle \left\vert 1\right\rangle  \label{eq2}
\end{eqnarray}%
The thing to notice is that although the superposition is made up of two
maximally entangled Bell basis, the results are not shift invariant (i.e.,
generation of another \textit{Wannier} state yields \textit{Wannier} state
located at another \textit{site, which is }orthogonal to the unshifted one).
This means that the superposition given above yields unentangled states
(product states) in Eq. (\ref{eq2}). In what follows, we will see that only
the following combination of entangled basis states yields another entangled
basis states as long as they both belong to either 'even' or 'odd' spaces,
i.e., by diagrammatic construction, we have,%
\begin{eqnarray}
triplet^{\pm }\otimes triplet^{\pm } &=&triplet^{\pm }\text{ or }%
singlet^{\pm }  \label{eq3} \\
triplet^{\pm }\otimes singlet^{\pm } &=&singlet^{\pm }\text{ or }%
triplet^{\pm }  \label{eq4} \\
singlet^{\pm }\otimes singlet^{\pm } &=&triplet^{\pm }\text{ or }%
singlet^{\pm }  \label{q5}
\end{eqnarray}

Equation (\ref{eq4}) is quite interesting because it holds on a complete
expansion of a direct product of two qubit states. These relations can
easily be deduced from the physical diagrammatic model, see Fig. \ref{fig1}.
For example, if we define the operation as a superposition such as 
\begin{eqnarray}
\frac{1}{\sqrt{2}}\left( \Phi ^{+}+\Psi ^{+}\right) &=&\frac{1}{\sqrt{2}}%
\left( \frac{1}{\sqrt{2}}\left( \left\vert 1\right\rangle \left\vert
1\right\rangle +\left\vert 0\right\rangle \left\vert 0\right\rangle \right) +%
\frac{1}{\sqrt{2}}\left( \left\vert 0\right\rangle \left\vert 1\right\rangle
+\left\vert 1\right\rangle \left\vert 0\right\rangle \right) \right)  \notag
\\
&=&\frac{1}{2}\left[ \left( \left\vert 1\right\rangle \left\vert
1\right\rangle +\left\vert 0\right\rangle \left\vert 0\right\rangle \right)
+\left( \left\vert 0\right\rangle \left\vert 1\right\rangle +\left\vert
1\right\rangle \left\vert 0\right\rangle \right) \right]  \notag \\
&=&\left[ \frac{1}{\sqrt{2}}\left( \left\vert 1\right\rangle +\left\vert
0\right\rangle \right) \otimes \frac{1}{\sqrt{2}}\left( \left\vert
1\right\rangle +\left\vert 0\right\rangle \right) \right]  \label{[prductent}
\end{eqnarray}%
We see that Eq. (\ref{[prductent}) is a direct product state of two qubits.
We will see in what follows that this is an entangled state and corresponds
to Eq. (\ref{eq4}) of the $singlet^{+}$ or $triplet^{+}$ of the model
diagrams depending on the actual linkage. Similarly, we have, 
\begin{eqnarray*}
\frac{1}{\sqrt{2}}\left( \Phi ^{-}+\Psi ^{-}\right) &=&\frac{1}{\sqrt{2}}%
\left( \frac{1}{\sqrt{2}}\left( \left\vert 1\right\rangle \left\vert
1\right\rangle -\left\vert 0\right\rangle \left\vert 0\right\rangle \right) +%
\frac{1}{\sqrt{2}}\left( \left\vert 0\right\rangle \left\vert 1\right\rangle
-\left\vert 1\right\rangle \left\vert 0\right\rangle \right) \right) \\
&=&\frac{1}{2}\left[ \left( \left\vert 1\right\rangle \left\vert
1\right\rangle -\left\vert 0\right\rangle \left\vert 0\right\rangle \right)
+\left( \left\vert 0\right\rangle \left\vert 1\right\rangle -\left\vert
1\right\rangle \left\vert 0\right\rangle \right) \right] \\
&=&\left[ \frac{1}{\sqrt{2}}\left( \left\vert 1\right\rangle +\left\vert
0\right\rangle \right) \otimes \frac{1}{\sqrt{2}}\left( \left\vert
1\right\rangle -\left\vert 0\right\rangle \right) \right]
\end{eqnarray*}%
corresponds to Eq. (\ref{eq4}) of the $singlet$ or $triplet$ of the model
diagrams, depending on the actual linking. \ 

However, the following combinations of 'even' and 'odd' entangled states
result in \textit{unentangled} states, namely,%
\begin{eqnarray*}
\frac{1}{\sqrt{2}}\left( \Phi ^{+}+\Psi ^{-}\right) &=&\frac{1}{\sqrt{2}}%
\left( \frac{1}{\sqrt{2}}\left( \left\vert 1\right\rangle \left\vert
1\right\rangle +\left\vert 0\right\rangle \left\vert 0\right\rangle \right) +%
\frac{1}{\sqrt{2}}\left( \left\vert 0\right\rangle \left\vert 1\right\rangle
-\left\vert 1\right\rangle \left\vert 0\right\rangle \right) \right) \\
&=&\frac{1}{2}\left[ \left( \left\vert 1\right\rangle \left\vert
1\right\rangle +\left\vert 0\right\rangle \left\vert 0\right\rangle \right)
+\left( \left\vert 0\right\rangle \left\vert 1\right\rangle -\left\vert
1\right\rangle \left\vert 0\right\rangle \right) \right]
\end{eqnarray*}%
and 
\begin{eqnarray*}
\frac{1}{\sqrt{2}}\left( \Phi ^{-}+\Psi ^{+}\right) &=&\frac{1}{\sqrt{2}}%
\left( \frac{1}{\sqrt{2}}\left( \left\vert 1\right\rangle \left\vert
1\right\rangle -\left\vert 0\right\rangle \left\vert 0\right\rangle \right) +%
\frac{1}{\sqrt{2}}\left( \left\vert 0\right\rangle \left\vert 1\right\rangle
+\left\vert 1\right\rangle \left\vert 0\right\rangle \right) \right) \\
&=&\frac{1}{2}\left[ \left( \left\vert 1\right\rangle \left\vert
1\right\rangle -\left\vert 0\right\rangle \left\vert 0\right\rangle \right)
+\left( \left\vert 0\right\rangle \left\vert 1\right\rangle +\left\vert
1\right\rangle \left\vert 0\right\rangle \right) \right]
\end{eqnarray*}%
by virtue of the failure to have global sign factors. All these claims are
justified through the concept of concurrence, an inherent property of
entangled qubits, to be discussed in what follows. Moreover, this feature of
failing to have global sign factor is also reflected in the failure to
represent by our inverter-chain link diagrams.

\section*{Bipartite System}

Let the Hilbert spaces of a bipartite system consisting of $A$ and $B$ be
denoted by $\mathcal{H}_{A}$ and $\mathcal{H}_{B}$, respectively. A
bipartite system is a system with Hilbert space equal to the direct product
of $\mathcal{H}_{A}$ and $\mathcal{H}_{B}$, i.e.,%
\begin{equation*}
\mathcal{H}_{AB}=\mathcal{H}_{A}\otimes \mathcal{H}_{B}
\end{equation*}%
Let the density matrix for the whole system be denoted by $\rho $. Then the
reduced density matrix of a subsystem $A$ is given by the partial trace,%
\begin{equation*}
\rho _{A}=Tr_{_{B}}\ \rho
\end{equation*}%
The entanglement entropy, $S_{A}$, is the \textit{von Newmann entropy} of
the reduced density matrix, $\rho _{A}$,%
\begin{equation*}
S_{A}=-Tr\ \rho _{A}\log \rho _{A}
\end{equation*}

\begin{example}
Let $\Omega $ be the number of distinct states in subsystem $A.$ Assume a
uniform distribution among states, hence $\rho _{A}$ has eigenvalues $\frac{1%
}{\Omega }$, i.e., $\rho _{A}$ can be represented by a diagonal $\Omega
\times \Omega $ matrix with identical matrix elements given by $\frac{1}{%
\Omega }$. Thus, in taking the trace we can use the eigenvalues of the
reduced density matrix operator, $\rho _{A}$. Therefore, we have%
\begin{eqnarray*}
S_{A} &=&-Tr\frac{1}{\Omega }\log \frac{1}{\Omega } \\
&=&-\Omega \left( \frac{1}{\Omega }\log \frac{1}{\Omega }\right) \\
&=&-\log \frac{1}{\Omega } \\
&=&\log \ \Omega
\end{eqnarray*}%
Upon multiplying by the Boltzmann constant, $k_{B}$, we obtain%
\begin{equation*}
k_{B}S_{A}=k_{B}\log \ \Omega
\end{equation*}%
which is the Boltzmann thermodynamic entropy, based on ergodic theorem.
\end{example}

\begin{example}
Two qubit system.
\end{example}

A qubit is simply a quantum bit whose number of distinct eigenstates is $2$.
We denote these eigenstates as $\left\vert 0\right\rangle $ and $\left\vert
1\right\rangle $, i.e., a two-state system. If each subsystem $A$ or $B$ is
a single qubit, then the Hilbert space of the whole system is span by the
following $4$ direct product states, namely,%
\begin{equation*}
\left\vert 00\right\rangle \text{, }\left\vert 01\right\rangle \text{, }%
\left\vert 10\right\rangle \text{, }\left\vert 11\right\rangle
\end{equation*}%
Here, the first bit refers to subsystem $A$ and the second bit refers to
subsystem $B$, where we write%
\begin{equation*}
\left\vert ij\right\rangle =\left\vert i\right\rangle _{A}\ \left\vert
j\right\rangle _{B}\ \epsilon \ \mathcal{H}_{A}\otimes \mathcal{H}_{B}
\end{equation*}%
Now, the density matrix of a pure state is,%
\begin{equation*}
\rho =\left\vert \psi \right\rangle \left\langle \psi \right\vert
\end{equation*}%
then%
\begin{eqnarray*}
\rho ^{2} &=&\left\vert \psi \right\rangle \left\vert \psi \right\rangle
\left\langle \psi \right\vert \left\vert \psi \right\rangle \left\langle
\psi \right\vert \\
&=&\left\vert \psi \right\rangle \left\langle \psi \right\vert \\
&=&\rho
\end{eqnarray*}%
An operator whose square is equal to itself must have an eigenvalue equal to
unity. \ Let us write for pure state of the two qubits as,%
\begin{eqnarray*}
\left\vert \psi \right\rangle &=&\frac{1}{\sqrt{2}}\left( \left\vert
00\right\rangle +\left\vert 11\right\rangle \right) \\
\left\langle \psi \right\vert &=&\frac{1}{\sqrt{2}}\left( \left\langle
00\right\vert +\left\langle 11\right\vert \right) \\
\left\vert \psi \right\rangle \left\langle \psi \right\vert &=&\frac{1}{2}%
\left( \left\vert 00\right\rangle +\left\vert 11\right\rangle \right) \left(
\left\langle 00\right\vert +\left\langle 11\right\vert \right)
\end{eqnarray*}%
We have%
\begin{eqnarray*}
\left\langle \psi \right\vert \left\vert \psi \right\rangle &=&\frac{1}{2}%
\left( \left\langle 00\right\vert +\left\langle 11\right\vert \right) \left(
\left\vert 00\right\rangle +\left\vert 11\right\rangle \right) \\
&=&\frac{1}{2}\left( \left\langle 00\right\vert \left\vert 00\right\rangle
+\left\langle 11\right\vert \left\vert 11\right\rangle \right) \\
&=&1
\end{eqnarray*}%
Thus, indeed,%
\begin{eqnarray*}
\rho ^{2} &=&\left\vert \psi \right\rangle \left\langle \psi \right\vert
\left\vert \psi \right\rangle \left\langle \psi \right\vert \\
&=&\left\vert \psi \right\rangle \left\langle \psi \right\vert \\
&=&\rho
\end{eqnarray*}%
We refer to $\left\vert \psi \right\rangle $ as a maximally entangled state
in the sense that the first qubit is exactly "mapped" to the second qubit.
The reduced density matrix for subsystem $A$ is%
\begin{eqnarray*}
\rho _{A} &=&Tr_{B}\ \rho \\
&=&\frac{1}{2}\left\langle 0_{B}\right\vert \left( \left\vert
00\right\rangle +\left\vert 11\right\rangle \right) \left( \left\langle
00\right\vert +\left\langle 11\right\vert \right) \left\vert
0_{B}\right\rangle \\
&&+\frac{1}{2}\left\langle 1_{B}\right\vert \left( \left\vert
00\right\rangle +\left\vert 11\right\rangle \right) \left( \left\langle
00\right\vert +\left\langle 11\right\vert \right) \left\vert
1_{B}\right\rangle \\
&=&\frac{1}{2}\left( \left\vert 0_{A}\right\rangle \left\langle
0_{A}\right\vert +\left\vert 1_{A}\right\rangle \left\langle
1_{A}\right\vert \right)
\end{eqnarray*}%
Now clearly%
\begin{eqnarray*}
\rho _{A}^{2} &=&\frac{1}{4}\left( \left\vert 0_{A}\right\rangle
\left\langle 0_{A}\right\vert +\left\vert 1_{A}\right\rangle \left\langle
1_{A}\right\vert \right) \left( \left\vert 0_{A}\right\rangle \left\langle
0_{A}\right\vert +\left\vert 1_{A}\right\rangle \left\langle
1_{A}\right\vert \right) \\
&=&\frac{1}{4}\left( \left\vert 0_{A}\right\rangle \left\langle
0_{A}\right\vert +\left\vert 1_{A}\right\rangle \left\langle
1_{A}\right\vert \right) \\
&=&\frac{1}{2}\rho _{A}
\end{eqnarray*}%
so that $\rho _{A}$ is not a pure state but mixed, i.e., a mixture of two
pure states, $\left\vert 0_{A}\right\rangle $ and $\left\vert
1_{A}\right\rangle $. Note that the mixed states density matrix do not
possess \textit{off-diagonal elements}. From the last identity, the
eigenvalues of $\rho _{A}$ is $\frac{1}{2}$ and since the subsystem $A$ is a
single qubit with two distinct states, then eigenvalues of $\rho _{A}$
corresponds to $\frac{1}{\Omega }$ of our first example above. Thus we refer
to $\rho _{A}$ as "uniformly'" mixed often referred to as "\textit{maximally
mixed}" with the initial state $\left\vert \psi \right\rangle $ as "\textit{%
maximally} \textit{entangled"}.

\section{Entanglement of Formation of Multi-Partite Qubit Systems}

The entanglement entropy of subsystem $A$ can be calculated using the
eigenvalues of the $2\times 2$ matrix of $\rho _{A}$, which is $\lambda _{A}=%
\frac{1}{2}$. Therefore we have for the entanglement entropy or entanglement
of formation given by,%
\begin{eqnarray}
S_{A} &=&-Tr\rho _{A}\log \rho _{A}  \notag \\
&=&-Tr\frac{1}{2}\log \frac{1}{2}  \notag \\
&=&-2\left( \frac{1}{2}\log \frac{1}{2}\right)  \notag \\
&=&\log 2\equiv \log 2^{1}  \label{eq6}
\end{eqnarray}%
where exponent base $2$, here $1$, is the number of qubits that is entangled
with system $B$. This will be made clear in the next example.

\begin{example}
A four qubit system:

If each subsystem $A$ or $B$ has two qubits, then the Hilbert space of the
whole system is span by the following $16$ direct product states,i.e., $%
2^{4}=16$ states. The maximally entangled pure state is determined by the
following eight diagrams \cite{mechanical}, see Fig.\ref{fig2}, and their
flipped or translated states,

\begin{figure}[h]
\centering
\includegraphics[width=3.4091in]{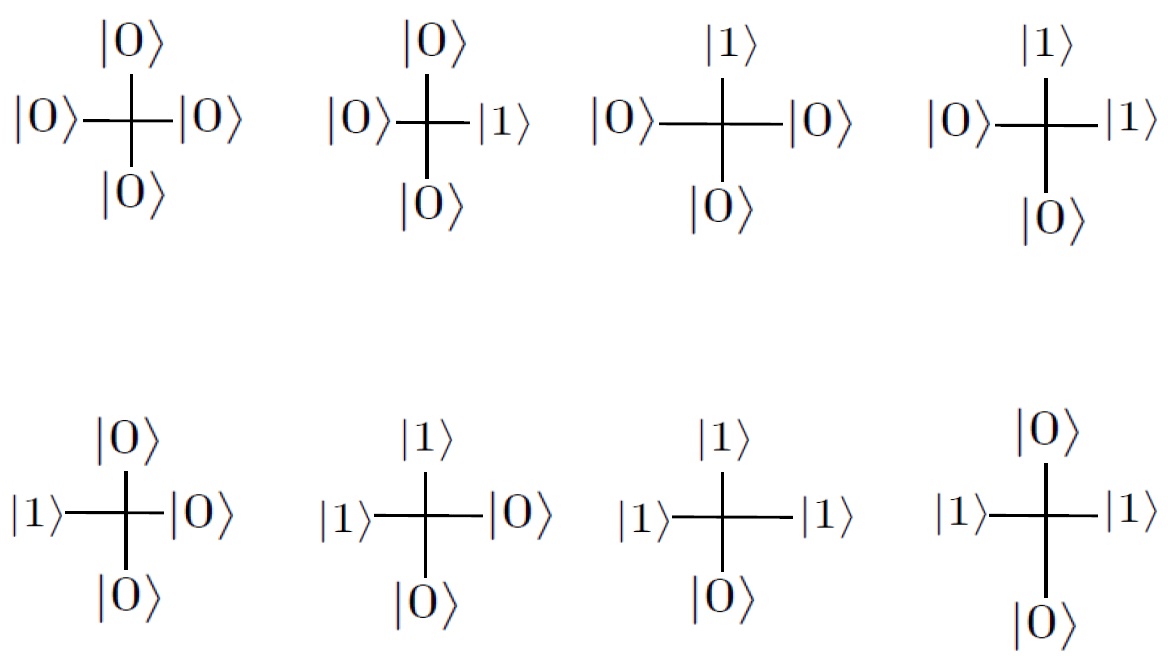}
\caption{Two-state eight diagrams for entangled four qubits. The so-called
flip operation yields the second state for each of the above diagrams. The
entangled basis is constructed by the superposition, via the Hadamard
transformation, of each diagram and its corresponding flipped diagram.}
\label{fig2}
\end{figure}

Any combination or superposition of both 'even' triplet and singlet states
comprised a maximally entangled state, i.e., 
\begin{equation*}
\left( 
\begin{array}{c}
\Phi _{1}^{+} \\ 
\Phi _{1}^{-}%
\end{array}%
\right) =\frac{1}{\sqrt{2}}\left( 
\begin{array}{cc}
1 & 1 \\ 
1 & -1%
\end{array}%
\right) \left( 
\begin{array}{c}
\left\vert 0\right\rangle \left\vert 0\right\rangle \left\vert
0\right\rangle \left\vert 0\right\rangle \\ 
\left\vert 1\right\rangle \left\vert 1\right\rangle \left\vert
1\right\rangle \left\vert 1\right\rangle%
\end{array}%
\right)
\end{equation*}%
\begin{equation*}
\left( 
\begin{array}{c}
\Phi _{8}^{+} \\ 
\Phi _{8}^{-}%
\end{array}%
\right) =\frac{1}{\sqrt{2}}\left( 
\begin{array}{cc}
1 & 1 \\ 
1 & -1%
\end{array}%
\right) \left( 
\begin{array}{c}
\left\vert 0\right\rangle \left\vert 1\right\rangle \left\vert
0\right\rangle \left\vert 1\right\rangle \\ 
\left\vert 1\right\rangle \left\vert 0\right\rangle \left\vert
1\right\rangle \left\vert 0\right\rangle%
\end{array}%
\right)
\end{equation*}%
Thus, from the states given above, we have for example the maximally
entangled state,%
\begin{eqnarray}
\left\vert \psi \right\rangle &=&\frac{1}{\sqrt{4}}\left( \left\vert
00,00\right\rangle +\left\vert 01,01\right\rangle +\left\vert
10,10\right\rangle +\left\vert 11,11\right\rangle \right)  \notag \\
&=&\frac{1}{\sqrt{2}}\left( \Phi _{1}^{+}+\Phi _{8}^{+}\right)  \label{eq7}
\end{eqnarray}%
and similarly, we can also form another entangled state,%
\begin{eqnarray}
\left\vert \psi \right\rangle &=&\frac{1}{\sqrt{4}}\left( \left\vert
00,00\right\rangle +\left\vert 01,01\right\rangle -\left\vert
10,10\right\rangle -\left\vert 11,11\right\rangle \right)  \notag \\
&=&\frac{1}{\sqrt{2}}\left( \Phi _{1}^{-}+\Phi _{8}^{-}\right)  \label{eq8}
\end{eqnarray}%
where the first two bits belongs to subsystem $A$ and the second two bits
belong to subsystem $B$. We note that%
\begin{eqnarray*}
\left\langle \psi \right\vert \left\vert \psi \right\rangle &=&\frac{1}{4}%
\left( \left\langle 00,00\right\vert +\left\langle 01,01\right\vert
+\left\langle 10,10\right\vert +\left\langle 11,11\right\vert \right) \left(
\left\vert 00,00\right\rangle +\left\vert 01,01\right\rangle +\left\vert
10,10\right\rangle +\left\vert 11,11\right\rangle \right) \\
&=&1
\end{eqnarray*}%
Therefore%
\begin{eqnarray*}
\rho ^{2} &=&\left( \left\vert \psi \right\rangle \left\langle \psi
\right\vert \right) ^{2} \\
&=&\left\vert \psi \right\rangle \left\langle \psi \right\vert =\rho
\end{eqnarray*}%
So $\left\vert \psi \right\rangle $ is a pure state.

In general, the following maximally entangled state corresponds to a chain
of entangled basis states, namely,%
\begin{equation*}
\left\vert \psi ^{+}\right\rangle =\frac{1}{\sqrt{8}}\left( \Phi
_{1}^{+}+\Phi _{2}^{+}+\Phi _{3}^{+}+\Phi _{4}^{+}+\Phi _{5}^{+}+\Phi
_{6}^{+}+\Phi _{7}^{+}+\Phi _{8}^{+}\right)
\end{equation*}%
as well as%
\begin{equation*}
\left\vert \psi ^{-}\right\rangle =\frac{1}{\sqrt{8}}\left( \Phi
_{1}^{-}+\Phi _{2}^{-}+\Phi _{3}^{-}+\Phi _{4}^{-}+\Phi _{5}^{-}+\Phi
_{6}^{-}+\Phi _{7}^{-}+\Phi _{8}^{-}\right)
\end{equation*}%
of a four qubit system.

The density matrix operator for the whole $4$-qubit system can be written as,%
\begin{equation*}
\rho =\frac{1}{\sqrt{2}}\left( \left\vert \Phi _{1}^{-}+\Phi
_{8}^{-}\right\rangle \right) \frac{1}{\sqrt{2}}\left( \left\langle \Phi
_{1}^{-}+\Phi _{8}^{-}\right\vert \right)
\end{equation*}%
\begin{equation*}
\rho =\frac{1}{4}\left[ \left\vert 00,00\right\rangle +\left\vert
01,01\right\rangle +\left\vert 10,10\right\rangle +\left\vert
11,11\right\rangle \right] \left[ \left\langle 00,00\right\vert
+\left\langle 01,01\right\vert +\left\langle 10,10\right\vert +\left\langle
11,11\right\vert \right]
\end{equation*}%
The reduced density matrix operator for subsystem $A$ is again obtained by
taking partial trace with respect to subsystem $B$.%
\begin{eqnarray*}
\rho _{A} &=&Tr_{B}\ \rho \\
&=&\left\langle 00_{B}\right\vert \rho \left\vert 00_{B}\right\rangle
+\left\langle 01_{B}\right\vert \rho \left\vert 01_{B}\right\rangle
+\left\langle 10_{B}\right\vert \rho \left\vert 10_{B}\right\rangle
+\left\langle 11_{B}\right\vert \rho \left\vert 11_{B}\right\rangle \\
&=&\frac{1}{4}\left\vert 00_{A}\right\rangle \left\langle 00_{A}\right\vert
+\left\vert 01_{A}\right\rangle \left\langle 01_{A}\right\vert +\left\vert
10_{A}\right\rangle \left\langle 10_{A}\right\vert +\left\vert
11_{A}\right\rangle \left\langle 11_{A}\right\vert
\end{eqnarray*}%
To determine the eigenvalues for $\rho _{A}$, we take its square,%
\begin{eqnarray*}
\rho _{A}^{2} &=&\frac{1}{16}\left( \left\vert 00_{A}\right\rangle
\left\langle 00_{A}\right\vert +\left\vert 01_{A}\right\rangle \left\langle
01_{A}\right\vert +\left\vert 10_{A}\right\rangle \left\langle
10_{A}\right\vert +\left\vert 11_{A}\right\rangle \left\langle
11_{A}\right\vert \right) \\
&&\times \left( \left\vert 00_{A}\right\rangle \left\langle
00_{A}\right\vert +\left\vert 01_{A}\right\rangle \left\langle
01_{A}\right\vert +\left\vert 10_{A}\right\rangle \left\langle
10_{A}\right\vert +\left\vert 11_{A}\right\rangle \left\langle
11_{A}\right\vert \right) \\
&=&\frac{1}{16}\left( \left\vert 00_{A}\right\rangle \left\langle
00_{A}\right\vert +\left\vert 01_{A}\right\rangle \left\langle
01_{A}\right\vert +\left\vert 10_{A}\right\rangle \left\langle
10_{A}\right\vert +\left\vert 11_{A}\right\rangle \left\langle
11_{A}\right\vert \right) \\
&=&\frac{1}{16}\rho _{A}
\end{eqnarray*}%
So we have%
\begin{equation*}
\rho _{A}^{2}=\frac{1}{16}\rho _{A}
\end{equation*}%
and the eigenvalues of $\rho _{A}$ is $\frac{1}{4}$. We can now calculate
the entanglement entropy of subsystem $A.$ We have%
\begin{eqnarray*}
S_{A} &=&-Tr\rho _{A}\log \rho _{A} \\
&=&-4\left[ \frac{1}{4}\log \frac{1}{4}\right] \\
&=&-\log \frac{1}{4} \\
&=&\log 2^{2}
\end{eqnarray*}%
The exponent $2$ correspond to the number of qubits that is entangled with
subsystem $B$. In general, for maximally entangled bipartite system $A$ and $%
B$, each having $k$ number of qubits $S_{A}$ is given by%
\begin{equation*}
S_{A}=\log 2^{k}
\end{equation*}%
Now of course, for this bipartite system, 
\begin{equation*}
S_{A}=S_{B}
\end{equation*}%
which simply means a complete matching of configurations of each system $A$
and $B$, respectively, otherwise some degrees of freedom will be hanging,
not matched or cannot be entangled.
\end{example}

\begin{example}
Tripartite sytem of three qubits:

The following diagrams represent the entangled tripartite system of qubits. 
\begin{figure}[h]
\centering
\includegraphics[width=1.5912in]{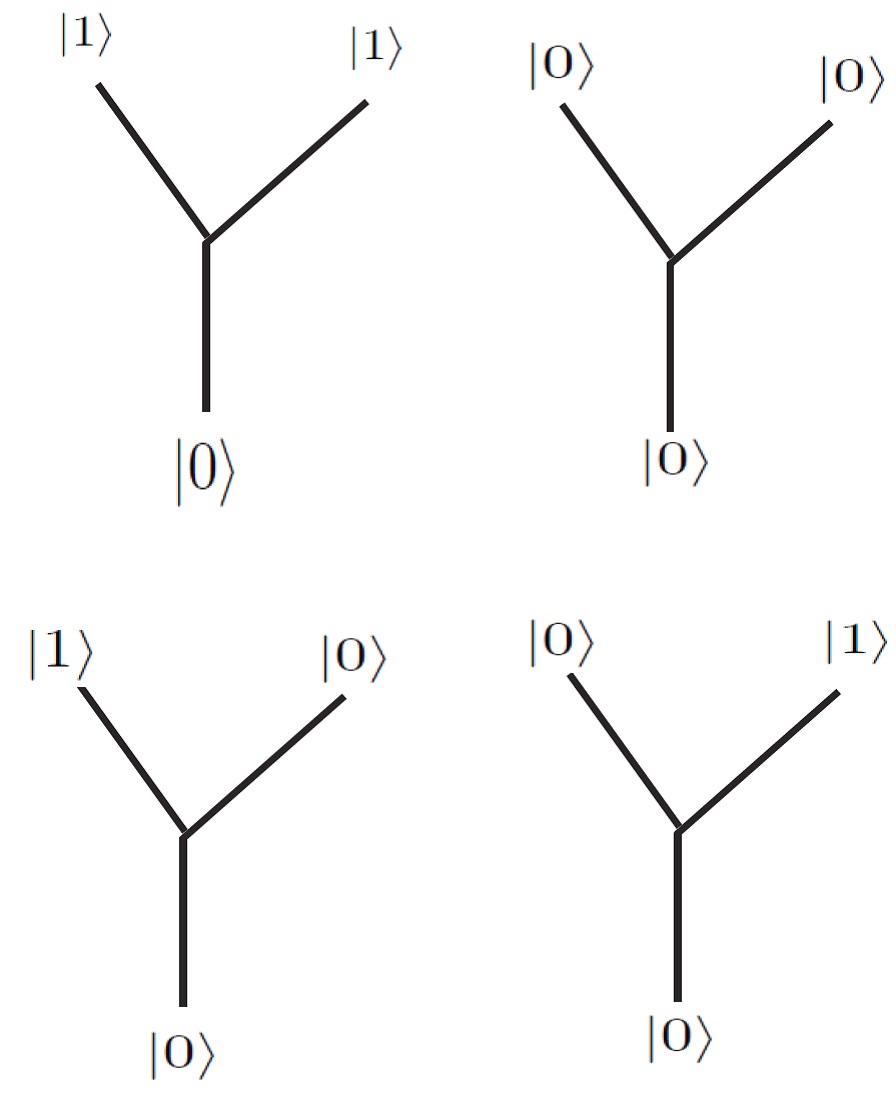}
\caption{Two-state four diagrams for entangled three qubits of a tripartite
system. Flip operations yield the respective second states.}
\label{fig9}
\end{figure}
The entangled basis states are as follows,%
\begin{equation*}
\left( 
\begin{array}{c}
\Xi ^{+} \\ 
\Xi ^{-}%
\end{array}%
\right) =\frac{1}{\sqrt{2}}\left( 
\begin{array}{cc}
1 & 1 \\ 
1 & -1%
\end{array}%
\right) \left( 
\begin{array}{c}
\left\vert 0\right\rangle \left\vert 1\right\rangle \left\vert 1\right\rangle
\\ 
\left\vert 1\right\rangle \left\vert 0\right\rangle \left\vert 0\right\rangle%
\end{array}%
\right)
\end{equation*}
\end{example}

\begin{equation*}
\left( 
\begin{array}{c}
\Theta _{3}^{+} \\ 
\Theta _{3}^{--}%
\end{array}%
\right) =\frac{1}{\sqrt{2}}\left( 
\begin{array}{cc}
1 & 1 \\ 
1 & -1%
\end{array}%
\right) \left( 
\begin{array}{c}
\left\vert 0\right\rangle \left\vert 0\right\rangle \left\vert 0\right\rangle
\\ 
\left\vert 1\right\rangle \left\vert 1\right\rangle \left\vert 1\right\rangle%
\end{array}%
\right)
\end{equation*}

\begin{equation*}
\left( 
\begin{array}{c}
\Omega ^{+} \\ 
\Omega ^{-}%
\end{array}%
\right) =\frac{1}{\sqrt{2}}\left( 
\begin{array}{cc}
1 & 1 \\ 
1 & -1%
\end{array}%
\right) \left( 
\begin{array}{c}
\left\vert 0\right\rangle \left\vert 1\right\rangle \left\vert 0\right\rangle
\\ 
\left\vert 1\right\rangle \left\vert 0\right\rangle \left\vert 1\right\rangle%
\end{array}%
\right)
\end{equation*}%
\begin{equation*}
\left( 
\begin{array}{c}
\Gamma ^{+} \\ 
\Gamma ^{-}%
\end{array}%
\right) =\frac{1}{\sqrt{2}}\left( 
\begin{array}{cc}
1 & 1 \\ 
1 & -1%
\end{array}%
\right) \left( 
\begin{array}{c}
\left\vert 0\right\rangle \left\vert 0\right\rangle \left\vert 1\right\rangle
\\ 
\left\vert 1\right\rangle \left\vert 1\right\rangle \left\vert 0\right\rangle%
\end{array}%
\right)
\end{equation*}%
A chain or superposition of any choice of entangled \textit{even} basis
states also form a maximally entangled state, e.g.,%
\begin{eqnarray*}
\left\vert \Psi ^{4}\right\rangle &=&\frac{1}{\sqrt{4}}\left[ \Xi
^{+}+\Theta _{3}^{+}+\Omega ^{+}+\Gamma ^{+}\right] \\
\left\vert \Psi ^{3}\right\rangle &=&\frac{1}{\sqrt{3}}\left[ \Xi
^{+}+\Theta _{3}^{+}+\Omega ^{+}\right] \\
\left\vert \Psi ^{2}\right\rangle &=&\frac{1}{\sqrt{2}}\left[ \Xi
^{+}+\Theta _{3}^{+}\right] \\
\left\vert \Psi ^{1}\right\rangle &=&\Xi ^{+}
\end{eqnarray*}%
are all, respectively, maximally entangled states with emergent two-state or
qubit properties.

\begin{example}
Here, we need to define what we mean by entanglement of formation as the
entanglement entropy of unentangling the three-qubit tripartite system into
three separate one-qubit systems. It is important to point out that any
entangled qubits, irrespective of their number, identically behave like a
single qubit, i.e., behaving exactly like a two-state system. Thus, the
natural order of disentangling is as follows: First, one entangles one qubit
from the remaining two entangled qubits. The entanglement of formation is
equivalent to one qubit. Next, one disentangled the remaining entangled two
qubits. This further give an entanglement of formation of one qubit. Thus,
the total entanglement of formation is $1+1=2$ qubits.
\end{example}

\subsection{On the monogamy inequality in multi-partite entanglement}

\begin{figure}[h]
\centering
\includegraphics[width=4.5143in]{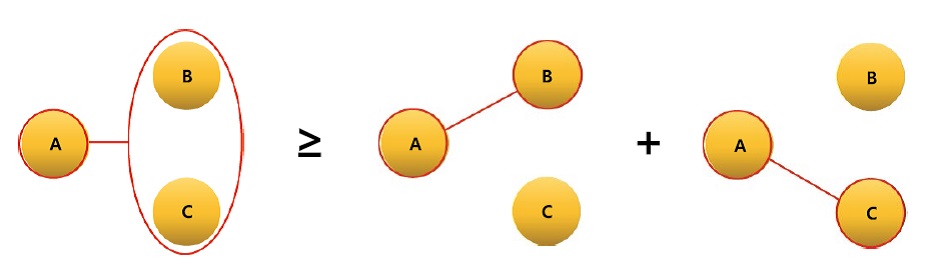}
\caption{Figure reproduced from Ref.\protect\cite{kim}. In our diagrammatic
analysis the right-hand side of the figure yields $1+1=2$ as the
entanglement of formation, being the sum of two two-qubit entanglements.
Thus, in our diagrammatic analysis, we obtained exact monogamy equality not
inequality.}
\label{fig10}
\end{figure}

The reasoning we have given above yields exact equality of the monogamy,
usually given as \textit{inequality} in the literature as deduced from
statistical analysis, e.g., one given by Kim \cite{kim}, The exact equality
relation for the entanglement of formation is diagrammatically shown in Fig. %
\ref{fig10}

The exact equality comes about by construction, since on both sides of Fig. %
\ref{fig10}, two $2$-qubit entangled states are being untangled to obtain
the entanglement of formation of this multi-partite system. This comes about
through the knowledge that any multi-partite entangled qubits behave as an
emergent qubit.

\begin{figure}[h]
\centering
\includegraphics[width=3.902in]{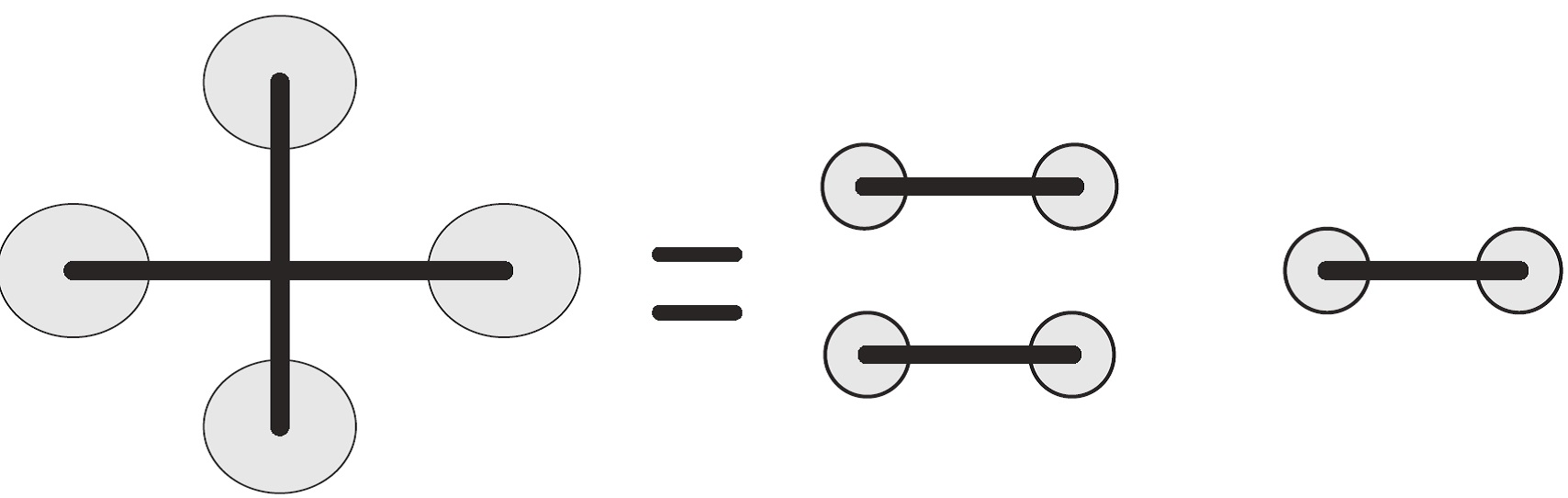}
\caption{In our diagrammatic analysis the right-hand side of the figure
yields $1+1+1=3$ as the entanglement of formation, being the sum of three
two-qubit entanglements. Thus, in our diagrammatic analysis, we obtained
exact equality not inequality.}
\label{fig11}
\end{figure}

\begin{figure}[h]
\centering
\includegraphics[width=3.4731in]{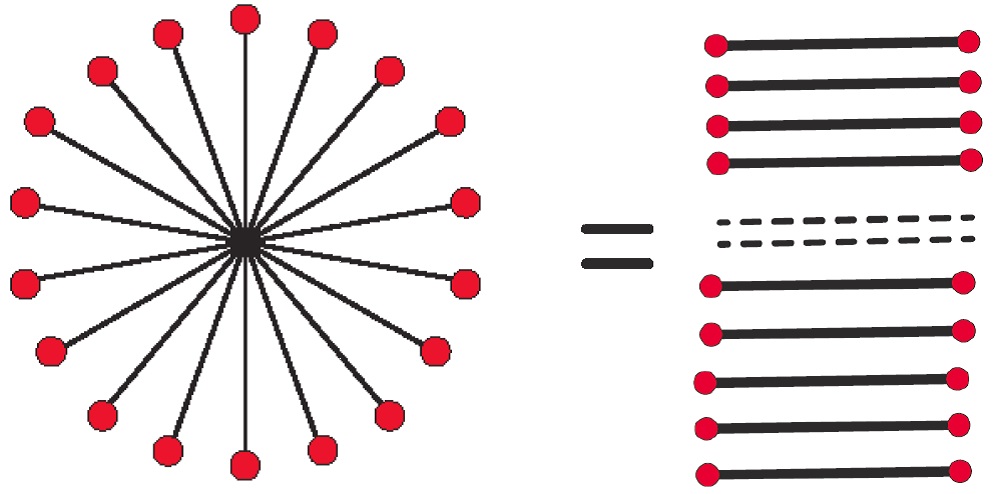}
\caption{In our diagrammatic analysis the right hand side of the figure
yields $1+1+1,......1=17$ qubits as the entanglement of formation, being the
sum of $17$ two-qubit entanglements of formation. There are $18$ multi-party
entangled qubits in the left-hand side.}
\label{fig12}
\end{figure}

Similarly, for an entangled $4$-partite system, the entanglement of
formation is determined schematically by the diagrams of Fig.\ref{fig11},
where the unentangling operations to be done on the left side are itemized
on the unentangling operations of \textit{three} entangled $2$-qubits on the
right side of the equality sign,

This sort of diagrammatic analysis of the entropy of entanglement of
formation can straightforwardly be employed to all multi-partite qubit
systems by a simple counting argument as is done here. Although our
diagrammatic analysis is based on maximally entangled qubits, we believe the
diagrammatic construction holds also for any arbitrarily entangled
multi-partite qubit systems. For example, for $18$ multi-partite entangled
qubits in Fig. \ref{fig12}, we have the entanglement of formation
schematically depicted by the equality where there is $17$ number of
monogamy in the right-hand side yielding $17$ qubits of entropy of
entanglement of formation.

\section{Expansion of Product States in Bell Basis}

Let us consider $\left\vert 00\right\rangle $, $\left\vert 01\right\rangle $%
, $\left\vert 10\right\rangle $, $\left\vert 11\right\rangle $ as our direct
product basis in expanding $\left\vert \psi \right\rangle $,%
\begin{eqnarray*}
\left\vert \psi \right\rangle &=&\alpha _{0}\left\vert 00\right\rangle
+\alpha _{1}\left\vert 01\right\rangle +\alpha _{2}\left\vert
10\right\rangle +\alpha _{3}\left\vert 11\right\rangle \\
&=&\sum\limits_{i}\beta _{i}e_{i}
\end{eqnarray*}%
where $e_{i}$ denotes the Bell basis. We have the Bell basis given by%
\begin{eqnarray}
e_{1} &=&\left\vert \Phi ^{+}\right\rangle =\frac{1}{\sqrt{2}}\left(
\left\vert 0\right\rangle \left\vert 0\right\rangle +\left\vert
1\right\rangle \left\vert 1\right\rangle \right)  \notag \\
e_{2} &=&\left\vert \Phi ^{-}\right\rangle =\frac{1}{\sqrt{2}}\left(
\left\vert 0\right\rangle \left\vert 0\right\rangle -\left\vert
1\right\rangle \left\vert 1\right\rangle \right)  \notag \\
e_{3} &=&\left\vert \Psi ^{+}\right\rangle =\frac{1}{\sqrt{2}}\left(
\left\vert 0\right\rangle \left\vert 1\right\rangle +\left\vert
1\right\rangle \left\vert 0\right\rangle \right)  \notag \\
e_{4} &=&\left\vert \Psi ^{-}\right\rangle =\frac{1}{\sqrt{2}}\left(
\left\vert 0\right\rangle \left\vert 1\right\rangle -\left\vert
1\right\rangle \left\vert 0\right\rangle \right)  \label{bijective}
\end{eqnarray}%
Standard basis can be express in terms of Bell basis, a re-statement of Eqs.
(\ref{eqa}) and (\ref{eqb}), namely,%
\begin{eqnarray*}
\left\vert 0\right\rangle \left\vert 0\right\rangle &=&\frac{1}{\sqrt{2}}%
\left( e_{1}+e_{2}\right) \\
\left\vert 1\right\rangle \left\vert 1\right\rangle &=&\frac{1}{\sqrt{2}}%
\left( e_{1}-e_{2}\right) \\
\left\vert 0\right\rangle \left\vert 1\right\rangle &=&\frac{1}{\sqrt{2}}%
\left( e_{3}+e_{4}\right) \\
\left\vert 1\right\rangle \left\vert 0\right\rangle &=&\frac{1}{\sqrt{2}}%
\left( e_{3}-e_{4}\right)
\end{eqnarray*}%
so we can also expand $\left\vert \psi \right\rangle $ as%
\begin{eqnarray*}
\left\vert \psi \right\rangle &=&\beta _{1}e_{1}+\beta _{2}e_{2}+\beta
_{3}e_{3}+\beta _{4}e_{4} \\
&=&\beta _{1}\left( \frac{1}{\sqrt{2}}\left( \left\vert 0\right\rangle
\left\vert 0\right\rangle +\left\vert 1\right\rangle \left\vert
1\right\rangle \right) \right) +\beta _{2}\left( \frac{1}{\sqrt{2}}\left(
\left\vert 0\right\rangle \left\vert 0\right\rangle -\left\vert
1\right\rangle \left\vert 1\right\rangle \right) \right) \\
&&+\beta _{3}\left( \frac{1}{\sqrt{2}}\left( \left\vert 0\right\rangle
\left\vert 1\right\rangle +\left\vert 1\right\rangle \left\vert
0\right\rangle \right) \right) +\beta _{4}\left( \frac{1}{\sqrt{2}}\left(
\left\vert 0\right\rangle \left\vert 1\right\rangle -\left\vert
1\right\rangle \left\vert 0\right\rangle \right) \right)
\end{eqnarray*}%
\begin{eqnarray*}
\left\vert \psi \right\rangle &=&\frac{1}{\sqrt{2}}\left( \beta _{1}+\beta
_{2}\right) \left\vert 0\right\rangle \left\vert 0\right\rangle +\frac{1}{%
\sqrt{2}}\left( \beta _{1}-\beta _{2}\right) \left\vert 1\right\rangle
\left\vert 1\right\rangle \\
&&+\frac{1}{\sqrt{2}}\left( \beta _{3}+\beta _{4}\right) \left\vert
0\right\rangle \left\vert 1\right\rangle +\frac{1}{\sqrt{2}}\left( \beta
_{3}-\beta _{4}\right) \left\vert 1\right\rangle \left\vert 0\right\rangle
\end{eqnarray*}%
Therefore, equating coefficients, we have the equality,%
\begin{eqnarray*}
\left\vert \psi \right\rangle &=&\alpha _{0}\left\vert 00\right\rangle
+\alpha _{1}\left\vert 01\right\rangle +\alpha _{2}\left\vert
10\right\rangle +\alpha _{3}\left\vert 11\right\rangle \\
&=&\frac{1}{\sqrt{2}}\left( \beta _{1}+\beta _{2}\right) \left\vert
0\right\rangle \left\vert 0\right\rangle +\frac{1}{\sqrt{2}}\left( \beta
_{1}-\beta _{2}\right) \left\vert 1\right\rangle \left\vert 1\right\rangle \\
&&+\frac{1}{\sqrt{2}}\left( \beta _{3}+\beta _{4}\right) \left\vert
0\right\rangle \left\vert 1\right\rangle +\frac{1}{\sqrt{2}}\left( \beta
_{3}-\beta _{4}\right) \left\vert 1\right\rangle \left\vert 0\right\rangle
\end{eqnarray*}%
yielding%
\begin{eqnarray*}
\alpha _{0} &=&\frac{1}{\sqrt{2}}\left( \beta _{1}+\beta _{2}\right) \\
\alpha _{1} &=&\frac{1}{\sqrt{2}}\left( \beta _{3}+\beta _{4}\right) \\
\alpha _{2} &=&\frac{1}{\sqrt{2}}\left( \beta _{3}-\beta _{4}\right) \\
\alpha _{3} &=&\frac{1}{\sqrt{2}}\left( \beta _{1}-\beta _{2}\right)
\end{eqnarray*}%
Then%
\begin{eqnarray*}
\sum\limits_{0}^{3}\left\vert \alpha _{i}\right\vert ^{2}
&=&\sum\limits_{1}^{4}\left\vert \beta _{i}\right\vert ^{2}=1 \\
&=&\left\vert \frac{1}{\sqrt{2}}\left( \beta _{1}+\beta _{2}\right)
\right\vert ^{2}+\left\vert \frac{1}{\sqrt{2}}\left( \beta _{1}-\beta
_{2}\right) \right\vert ^{2}+\left\vert \frac{1}{\sqrt{2}}\left( \beta
_{3}+\beta _{4}\right) \right\vert ^{2}+\left\vert \frac{1}{\sqrt{2}}\left(
\beta _{3}-\beta _{4}\right) \right\vert ^{2} \\
&=&\frac{1}{2}\left( 2\left( \beta _{1}^{2}+\beta _{2}^{2}\right) \right) +%
\frac{1}{2}\left( 2\left( \beta _{3}^{2}+\beta _{4}^{2}\right) \right)
=\sum\limits_{1}^{4}\left\vert \beta _{i}\right\vert ^{2}=1
\end{eqnarray*}%
Summarizing, the bijective relation between the Bell basis states and the
product states given in Eq. (\ref{bijective}) reflects the discrete
phase-space physics of these two-state systems.

\section{Chiral Degrees of Freedom and Entangled Qubits}

Observe that the pseudo-spin variables in semiconductor Bloch equation is
defined by the following expressions, 
\begin{eqnarray*}
S_{0} &=&\left( \rho _{cc}+\rho _{vv}\right) \\
S_{x} &=&\rho _{vc}+\rho _{cv} \\
S_{y} &=&-i\left( \rho _{vc}-\rho _{cv}\right) \\
S_{z} &=&\left( \rho _{cc}-\rho _{vv}\right)
\end{eqnarray*}%
This maps to 
\begin{eqnarray*}
e_{0} &=&\left\vert \Phi ^{+}\right\rangle =\frac{1}{\sqrt{2}}\left(
\left\vert 1\right\rangle \left\vert 1\right\rangle +\left\vert
0\right\rangle \left\vert 0\right\rangle \right) \\
e_{x} &=&\left\vert \Psi ^{+}\right\rangle =\frac{1}{\sqrt{2}}\left(
\left\vert 0\right\rangle \left\vert 1\right\rangle +\left\vert
1\right\rangle \left\vert 0\right\rangle \right) \\
ie_{y} &=&i\left\vert \Psi ^{-}\right\rangle =\frac{1}{\sqrt{2}}\left(
\left\vert 0\right\rangle \left\vert 1\right\rangle -\left\vert
1\right\rangle \left\vert 0\right\rangle \right) \\
e_{z} &=&\left\vert \Phi ^{-}\right\rangle =\frac{1}{\sqrt{2}}\left(
\left\vert 1\right\rangle \left\vert 1\right\rangle -\left\vert
0\right\rangle \left\vert 0\right\rangle \right)
\end{eqnarray*}

We refer to the basis $\left( e_{0},e_{x},e_{y},e_{z}\right) $ as the
pseudo-spin component basis, which differ from the Bell basis in $\left\vert
\Psi ^{-}\right\rangle $ by the factor $i$ in $\left\vert \Psi
^{-}\right\rangle $. An important observation that follows from this is that
the entangling of chiral degrees of freedom creates another chiral degrees
of freedom, e.g., Eq. (\ref{eq8}).

This is a very important observation. This will aid in understanding the 
\textit{entanglement-induced delocalization, }for example, in topological
insulators where entanglement occurs at the edge of the sample. This is
treated by the quantum transport approach in Ref.\cite{physica}.

\section{The Concurrence Concept and Emergent Qubit}

By virtue of our diagrammatic construction, coupled with discrete phase
space Hadamard transformation, in deriving the entangled basis states, the
concurrence concept defined by Wooters \cite{wooters1, wooters2}, as well as
the two-state (qubit) properties of entangled multi-partite qubits,
naturally coincides with the mathematical description of our physical model
of qubit entanglement. In other words, the essence of the quantum
description of an entangled qubits in Fig. \ref{fig1} consists of the
superposition of product 'site states' and their corresponding translated
'site states' of all qubits. This means a superposition of the two states of
the entangled two qubits of Fig.\ref{fig1}. One observe that irrespective of
how many qubits are entangled, the resulting entangled state is an \textit{%
emergent} two-state system and therefore behave just like one qubit with
two-states, namely, the first state and its \textit{translated} state \cite%
{mechanical}. Therefore its associated entropy is just one, Eq. (\ref{eq6}).%
\footnote{%
This is also deduced when concurrence $C=1$ \cite{wooters2}.}

The concept of concurrence is basically contained by construction of our
physical model of qubit entanglement. It is defined by%
\begin{equation}
\left\vert \left\langle \Psi \right\vert \left\vert \tilde{\Psi}%
\right\rangle \right\vert =C  \label{defC}
\end{equation}%
where $C$ is the quantitative value of concurrence. This number lies between 
$0$ and $1,$ $0$ $\leq C\leq $ $1$. Here $\tilde{\Psi}$ is the corresponding
translated qubits of $\Psi $.

\section{A Natural Measure of Entanglement}

The two properties of any entangled qubits, namely, concurrence and its 
\textit{emergent} qubit behavior, lead Wooters \cite{wooters1, wooters2} to
introduce a measure of entanglement, as incorporated in the two formulas,%
\begin{equation}
E\left( C\right) =H\left( \frac{1}{2}+\frac{1}{2}\sqrt{1-C^{2}}\right)
\label{concur}
\end{equation}%
\begin{equation}
H\left( x\right) =-x\ln x-\left( 1-x\right) \ln \left( 1-x\right)
\label{qubit}
\end{equation}%
Equation (\ref{concur}) basically say that if there is complete concurrence,
i.e., $C=1$, then Eq. (\ref{qubit}) says that the system behave as an
emergent qubit or two-state system, as depicted clearly in our diagrams.
Thus, for maximally entangled multi-partite qubits, we have $C=1$,%
\begin{eqnarray*}
E\left( C\right) &=&H\left( \frac{1}{2}\right) \\
H\left( \frac{1}{2}\right) &=&\frac{1}{2}\ln 2+\frac{1}{2}\ln 2 \\
&=&1
\end{eqnarray*}%
affirming that entangled qubits behave as a two-state system or as an 
\textit{emergent qubit}, yielding entropy equals one.

\subsection{Entropic distance in entanglement measure}

From the above developments, one can introduce an entanglement entropic
distance by the formula%
\begin{equation*}
\mathcal{E}=\left\vert Tr\left( \rho \ln \rho \right) -1\right\vert
\end{equation*}%
where $Tr\left( \rho \ln \rho \right) $ is evaluated using Eqs. (\ref{concur}%
) and (\ref{qubit}). This distance has a range: $0\leq \mathcal{E\leq }1.$
When the concurrence, $C=1$, $Tr\left( \rho \ln \rho \right) =1=H\left( 
\frac{1}{2}\right) $, then we have the entropy-distance from maximally
entangled state, $\mathcal{E}=0$. When $C=0$, $Tr\left( \rho \ln \rho
\right) =0=H\left( 1\right) $, then we have the entropy-distance from
maximally entangled state, $\mathcal{E}=1$.

An example where the $C<1$ may occur in the following singlet state, with
less entanglement,%
\begin{equation*}
\left\vert \psi \right\rangle =\alpha \left\vert 01\right\rangle +\beta
\left\vert 10\right\rangle
\end{equation*}%
where 
\begin{equation*}
\alpha ^{2}+\beta ^{2}=1
\end{equation*}%
Then%
\begin{eqnarray*}
C &=&\left\vert \left\langle \psi \right\vert \left\vert \tilde{\psi}%
\right\rangle \right\vert =\left( \alpha \left\langle 01\right\vert +\beta
\left\langle 10\right\vert \right) \left( \alpha \left\vert 10\right\rangle
+\beta \left\vert 01\right\rangle \right) \\
&=&\alpha \beta +\beta \alpha \leq 1
\end{eqnarray*}

\section{Mixed States and Pure States}

In discussing mixed and pure states, one makes use of density-matrix
operators. This is the domain of abstract statistical treatment usually
found in the literature, perhaps following the statistical tradition of
Bell's theorem. To elucidate the basic physics, we will here avoid abstract
statistical treatment and only discuss specific situations and examples.

\subsection{Mixed states and mixed entanglements}

Consider the maximally mixed state of a triplet system,%
\begin{equation*}
\hat{\rho}_{W}=\frac{1}{2}\left( \left\vert 00\right\rangle \left\langle
00\right\vert +\left\vert 11\right\rangle \left\langle 11\right\vert \right)
\end{equation*}%
From Eqs. (\ref{eqa}), we obtain the mixed entanglements given by%
\begin{equation*}
\hat{\rho}_{B}=\frac{1}{2}\left( \left\vert \Phi ^{+}\Phi ^{+}\right\rangle
\left\langle \Phi ^{+}\Phi ^{+}\right\vert +\left\vert \Phi ^{-}\Phi
^{-}\right\rangle \left\langle \Phi ^{-}\Phi ^{-}\right\vert \right)
\end{equation*}%
Similarly, consider the maximally mixed state of a singlet system,%
\begin{equation*}
\hat{\rho}_{W}=\frac{1}{2}\left( \left\vert 01\right\rangle \left\langle
01\right\vert +\left\vert 10\right\rangle \left\langle 10\right\vert \right)
\end{equation*}%
From Eqs. (\ref{eqb}), we obtain the mixed entanglements given by%
\begin{equation*}
\hat{\rho}_{B}=\frac{1}{2}\left( \left\vert \Psi ^{+}\Psi ^{+}\right\rangle
\left\langle \Psi ^{+}\Psi ^{+}\right\vert +\left\vert \Psi ^{-}\Psi
^{-}\right\rangle \left\langle \Psi ^{-}\Psi ^{-}\right\vert \right)
\end{equation*}

\subsection{Mixed state from pure state (entangled state)}

Consider the example of a bipartite of two qubit system. Consider a pure
entangled state,%
\begin{equation*}
\left\vert \psi \right\rangle =\frac{1}{\sqrt{2}}\left( \left\vert
00\right\rangle +\left\vert 11\right\rangle \right) .
\end{equation*}%
Then the density matrix is,%
\begin{equation}
\rho =\frac{1}{2}\left( \left\vert 00\right\rangle +\left\vert
11\right\rangle \right) \left( \left\langle 00\right\vert +\left\langle
11\right\vert \right) ,  \label{pure}
\end{equation}%
where the first qubit belongs to party $A$ and the second qubit belongs to
party $B$.\footnote{%
What we mean by party $A$ and $B$ is in the general sense since any
entangled number of qubits behave as an emergent qubit.} We have 
\begin{equation*}
\rho ^{2}=\rho
\end{equation*}%
so that Eq. (\ref{pure}) is a pure state. Again, we have, by tracing the
party $B$ we obtain,

\begin{eqnarray*}
\rho _{A} &=&Tr_{B}\ \rho \\
&=&\frac{1}{2}\left\langle 0_{B}\right\vert \left( \left\vert
00\right\rangle +\left\vert 11\right\rangle \right) \left( \left\langle
00\right\vert +\left\langle 11\right\vert \right) \left\vert
0_{B}\right\rangle \\
&&+\frac{1}{2}\left\langle 1_{B}\right\vert \left( \left\vert
00\right\rangle +\left\vert 11\right\rangle \right) \left( \left\langle
00\right\vert +\left\langle 11\right\vert \right) \left\vert
1_{B}\right\rangle \\
&=&\frac{1}{2}\left( \left\vert 0_{A}\right\rangle \left\langle
0_{A}\right\vert +\left\vert 1_{A}\right\rangle \left\langle
1_{A}\right\vert \right)
\end{eqnarray*}%
Now clearly%
\begin{eqnarray*}
\rho _{A}^{2} &=&\frac{1}{4}\left( \left\vert 0_{A}\right\rangle
\left\langle 0_{A}\right\vert +\left\vert 1_{A}\right\rangle \left\langle
1_{A}\right\vert \right) \left( \left\vert 0_{A}\right\rangle \left\langle
0_{A}\right\vert +\left\vert 1_{A}\right\rangle \left\langle
1_{A}\right\vert \right) \\
&=&\frac{1}{4}\left( \left\vert 0_{A}\right\rangle \left\langle
0_{A}\right\vert +\left\vert 1_{A}\right\rangle \left\langle
1_{A}\right\vert \right) \\
&=&\frac{1}{2}\rho _{A}
\end{eqnarray*}%
so that $\rho _{A}$ is not a pure state but mixed, i.e., a mixture of two
pure states, $\left\vert 0_{A}\right\rangle $ and $\left\vert
1_{A}\right\rangle $.\footnote{%
If the parties $A$ and $B$ are entangled states, then what we have obtain
are also mixed entanglement through the inverse relationships, Eq. (\ref{eqa}%
) and (\ref{eqb}).} Note that the mixed states density matrix do not possess 
\textit{off-diagonal elements}. The tracing operation basically eliminates
the contribution of the "\textit{off-diagonal}" terms. From the last
identity, the eigenvalues of $\rho _{A}$ is $\frac{1}{2}$ and since the
subsystem $A$ is an emergent qubit, in general with two distinct states,
then eigenvalues of $\rho _{A}$ corresponds to $\frac{1}{\Omega }$ of our
first example above. Thus we refer to $\rho _{A}$ as "uniformly'" mixed
often referred to as "\textit{maximally mixed}" with the inital state $%
\left\vert \psi \right\rangle $ as "\textit{maximally} \textit{entangled"}.

\section{Concluding Remarks}

The discrete phase-space viewpoint for qubit entanglement has been very
fruitful in analytically deriving the entanglement basis states for any
number of entangled qubits as demonstrated by the author in his book \cite%
{book}, as well as in Ref. \cite{mechanical}. Moreover, the inverter-chain
mechanical model of entanglement link has been demonstrated to faithfully
implement the discrete phase space viewpoint \cite{mechanical}. The crucial
observation that arise from this inverter-chain link model is that any
multi-partite qubit entangled system has the property of an \textit{emergent}
qubit, which has been rigorously justified. This aspect of the model readily
lead us to the equality relations of the entropy of entanglement formation
in the so-called monogamy inequality of entanglement formation, using
statistical arguments \cite{kim} discussed in the literature. In this paper,
mixed states and mixed entanglements are related by the Hadamard
transformations. In addition, we show that mixed state can be extracted from
pure entangled state, where the party $A$ or $B$ need not be a single qubit
themselves but can each be an entangled multi-partite qubit system,
respectively, by virtue of \textit{emergent} qubit property of entangled
qubit systems.

The natual measure of entanglement should be based on entropy of
entanglement formation, concurrence, and entropic distance from maximally
entangled reference.

\end{document}